\documentclass[12pt]{article}
\usepackage{epsfig}

\begin{document}

\title{Using the Singular Spectrum Analysis for Investigation of Troposphere Parameters}
\author{Natalia Miller, Zinovy Malkin}
\date{Pulkovo Observatory, St. Petersburg, Russia}
\maketitle

\begin{abstract}
In this paper, the method of Singular Spectrum Analysis (SSA)
is applied for investigation of the zenith troposphere delay time-series
derived from VLBI observations.  With the help of this method we can
analyze the structure of time-series and separate the harmonic
and irregular (trend) components. Combined IVS time-series of zenith
wet and total troposphere delays obtained in IGG were used for analysis.
For this study, several VLBI stations with the most long time series of
troposphere zenith delays were selected, also taking into consideration
the geographic region where the station is located.  The investigations
were carried out using SSA mode.  As a result, trends and seasonal
components (with annual and semiannual periods) were obtained for all
the stations.  Using of SSA enabled us to determine nonlinear trends in
zenith delay, and also to study variations in the amplitude and the phase
of the seasonal components with time.
\end{abstract}

\vfill
\noindent \hrule width 0.4\textwidth
~\vskip 0.2ex
\noindent {\small 5th IVS General Meeting, St.~Petersburg, Russia, 3--6 March 2007}
\eject

\section{SSA Method}
In this research, we have investigated the combined IVS troposphere
zenith delay (TZD) series and focused on behavior trends and seasonal
components with the help of Singular Spectrum Analysis (SSA)
\cite{Golyand}. Additional information on SSA method, its abilities
and the corresponding software can be found on the site
http://www.gistatgroup.com/cat/.
With the help of this method we can:

-- Recognize certain components in the equally spaced time series,
which have been obtained from observations. The result of such procedure
is decomposition of time series into components that usually can be
identified as trends, periodical or oscillatory and noise components;

-- Extract components with the well-known period and estimate value
of phase shift and variation of amplitude of pseudo-harmonic signals;

-- Find periodicities that are not known in advance;

-- Extract trends of different resolutions. Natural decomposition of the
time series is constructed on the base of the unique parameter (the window
length). Grouping different subsets of the decomposition components one can
obtain both the tendency and accurate trend;

-- On the basis of the chosen components smooth out the initial data.

Contrast to the standard spectral analysis, where the basic functions
are given a priory as the sines and cosines of the Fourier method,
in SSA they are determined from the very data to form orthogonal basis.

\section{Analysis of Zenith Delay Time Series}
\begin{figure}
\centering
\epsfclipon \epsfxsize=\textwidth \epsffile{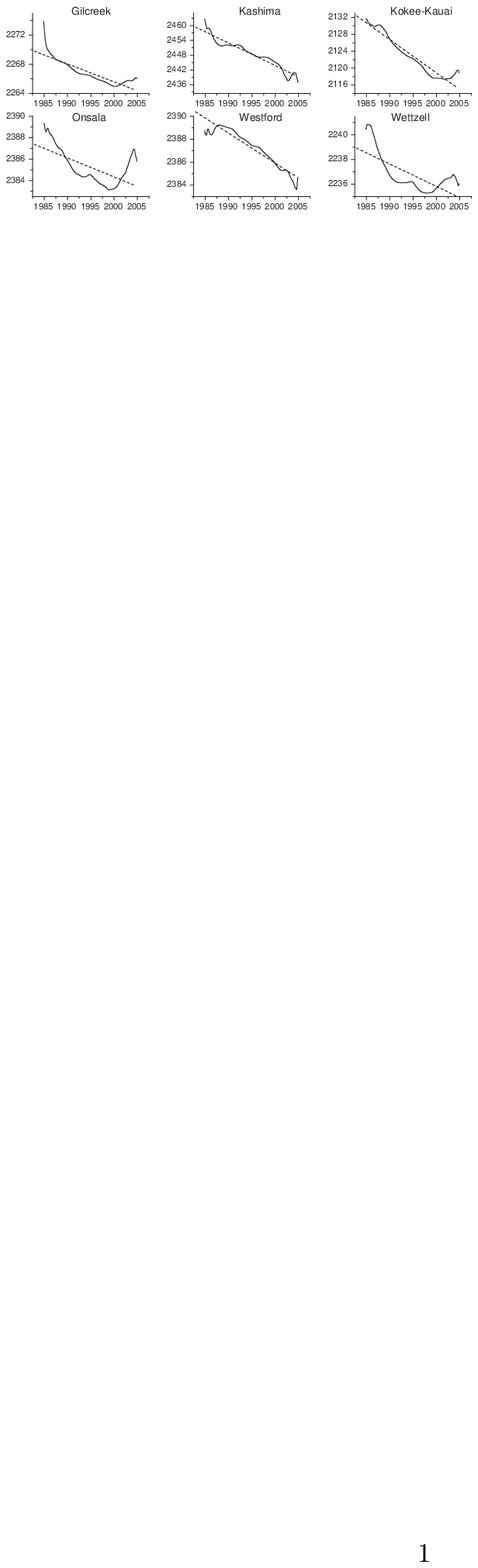}
\caption{SSA trends of ZTD ({\it solid line}) and their linear approximation
  ({\it dashed line}).  Unit: mm.}
\label{fig:trend1}
\end{figure}

\begin{figure}
\centering
\epsfclipon \epsfxsize=\textwidth \epsffile{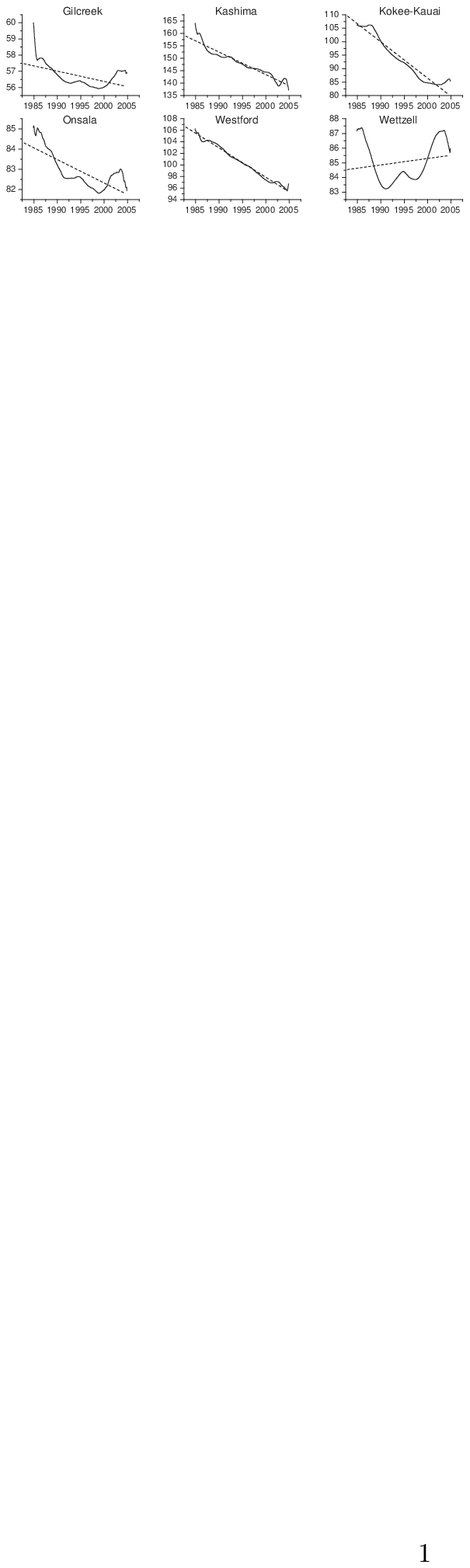}
\caption{SSA trends of ZTD ({\it solid line}) and their linear approximation
  ({\it dashed line}).  Unit: mm.}
\label{fig:trend2}
\end{figure}

\begin{figure}
\centering
\epsfclipon \epsfxsize=0.6\textwidth \epsffile{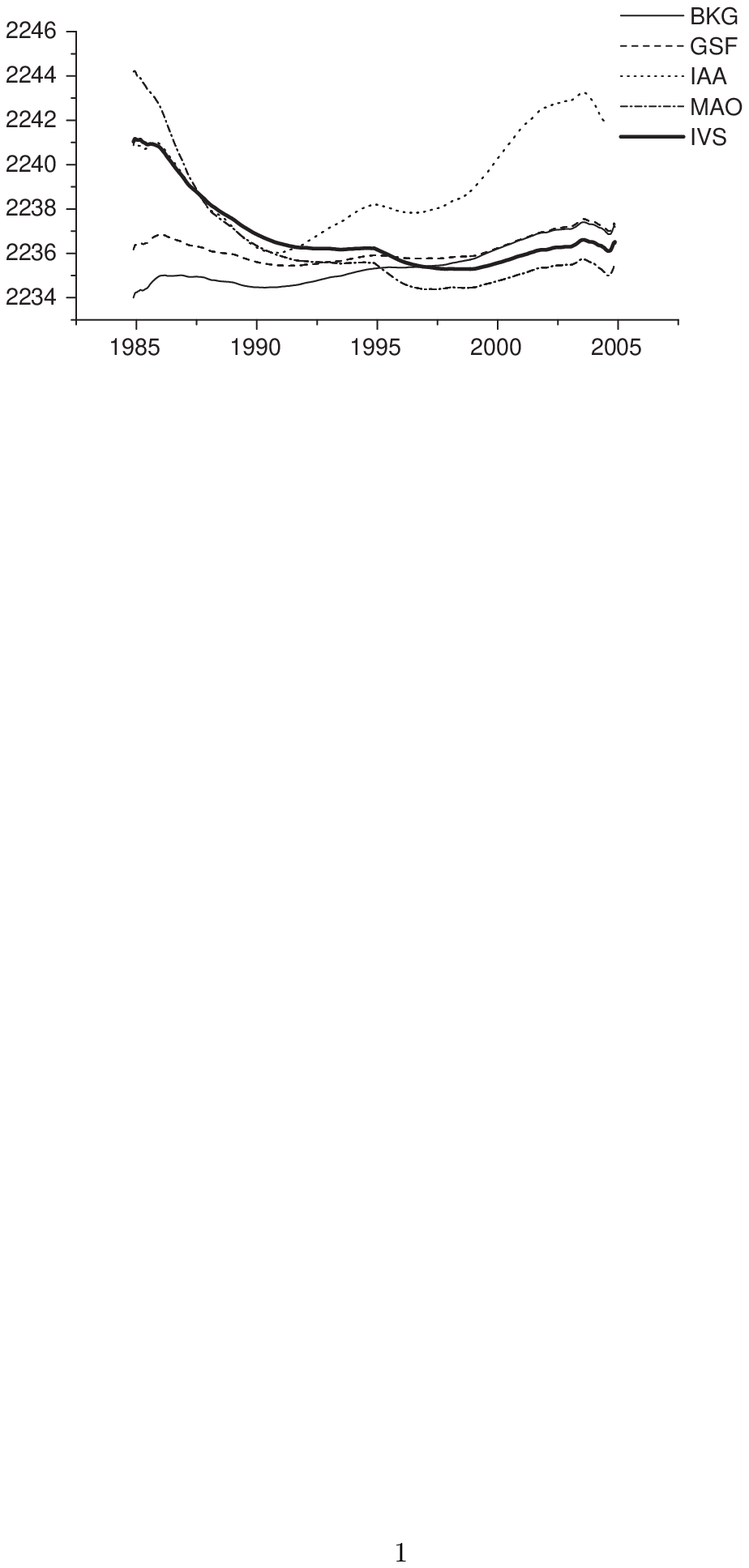}
\caption{SSA trends in ZTD for individual series for Wettzell. Unit: mm.}
\label{fig:trendWett}
\end{figure}

The zenith total delay (ZTD) is the sum of the zenith hydrostatic (ZHD)
and zenith wet delays (ZWD). The SSA method gives the opportunity to detect
features of main ZTD and ZWD (ZD) trends and to compare them with the main
trends of other time series, such as hydrostatic zenith delay,
wet zenith delay, pressure at the site, temperature at the site,
water vapor pressure at site, which were taken from the VMF1 files provided
by the IGG. Combined IVS time-series of ZD,obtained at IGG,
were used for analysis \cite{Heinkel}. Six VLBI stations (Gilcreek, Kashima,
Kokee-Kauai, Onsala, Westford, Wettzell) with the longest time series
of troposphere zenith delays were selected for studying.
Linear interpolation of data was carried out to get equally spaced
series with the step of 0.01 year which was used for SSA.
For all time series we used the same time interval 1984.88 -- 2004.87,
series length N=2000 points (20 years), and the maximum window length
M=N/2=1000. After decomposing with SSA method in all series the trends
hereafter referred to as trend SSA, annual and semiannual components were
found.
Figures~\ref{fig:trend1} and ~\ref{fig:trend2} show ZD trends as
obtained by SSA from the combined IVS series and their linear approximation.
The fact that the stations located in the same geographic region
(Onsala and Wettzell) reveal similar trend features is of a special interest.
Moreover, all trends have the same small curving about year 1995.
Original series obtained by analysis centers BKG, GSFC, IAA, MAO
show the same properties, which are shown in Fig.~\ref{fig:trendWett}
for Wettzell as an example.

\begin{figure}
\centering
\epsfclipon \epsfxsize=0.495\textwidth \epsffile{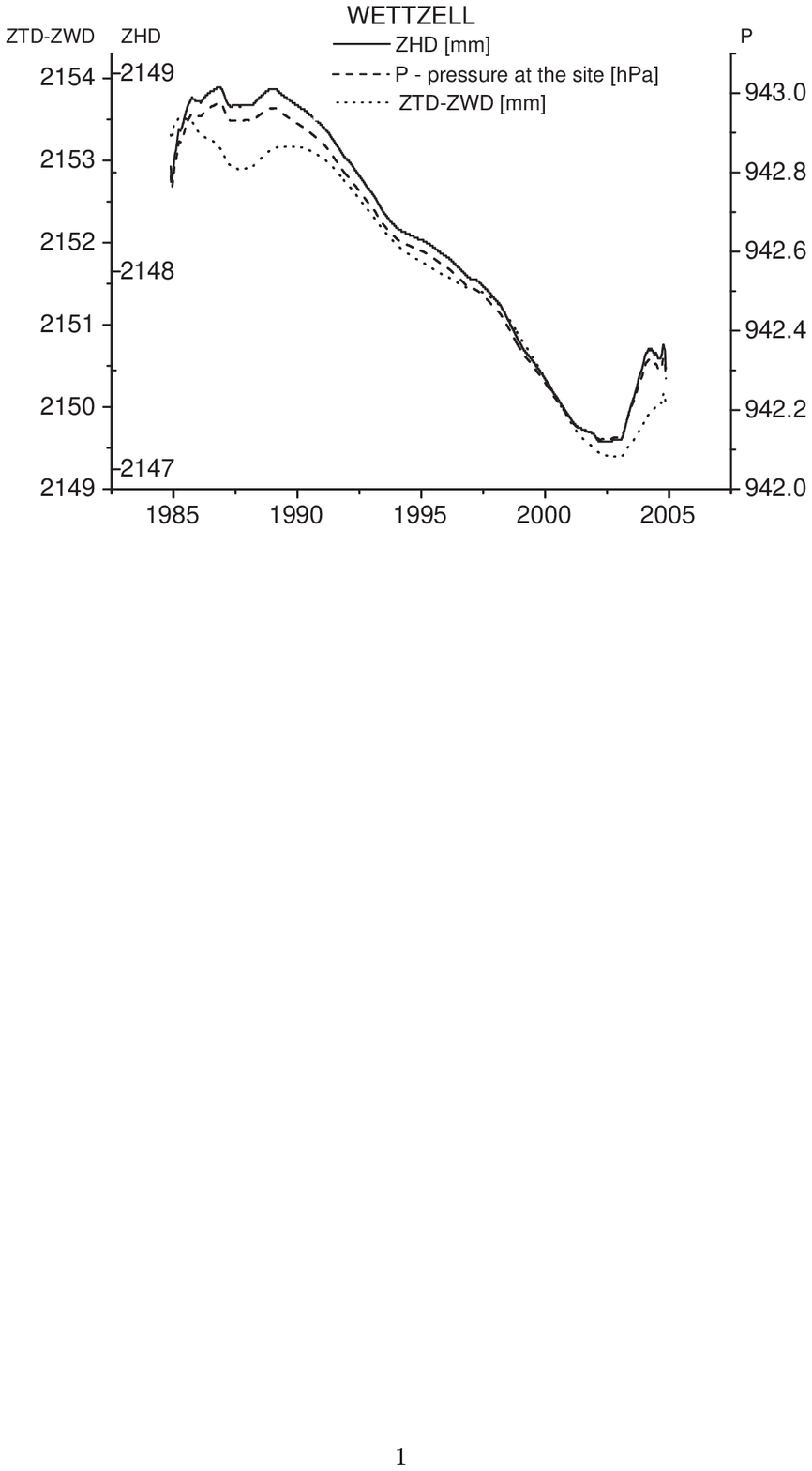}
\epsfclipon \epsfxsize=0.495\textwidth \epsffile{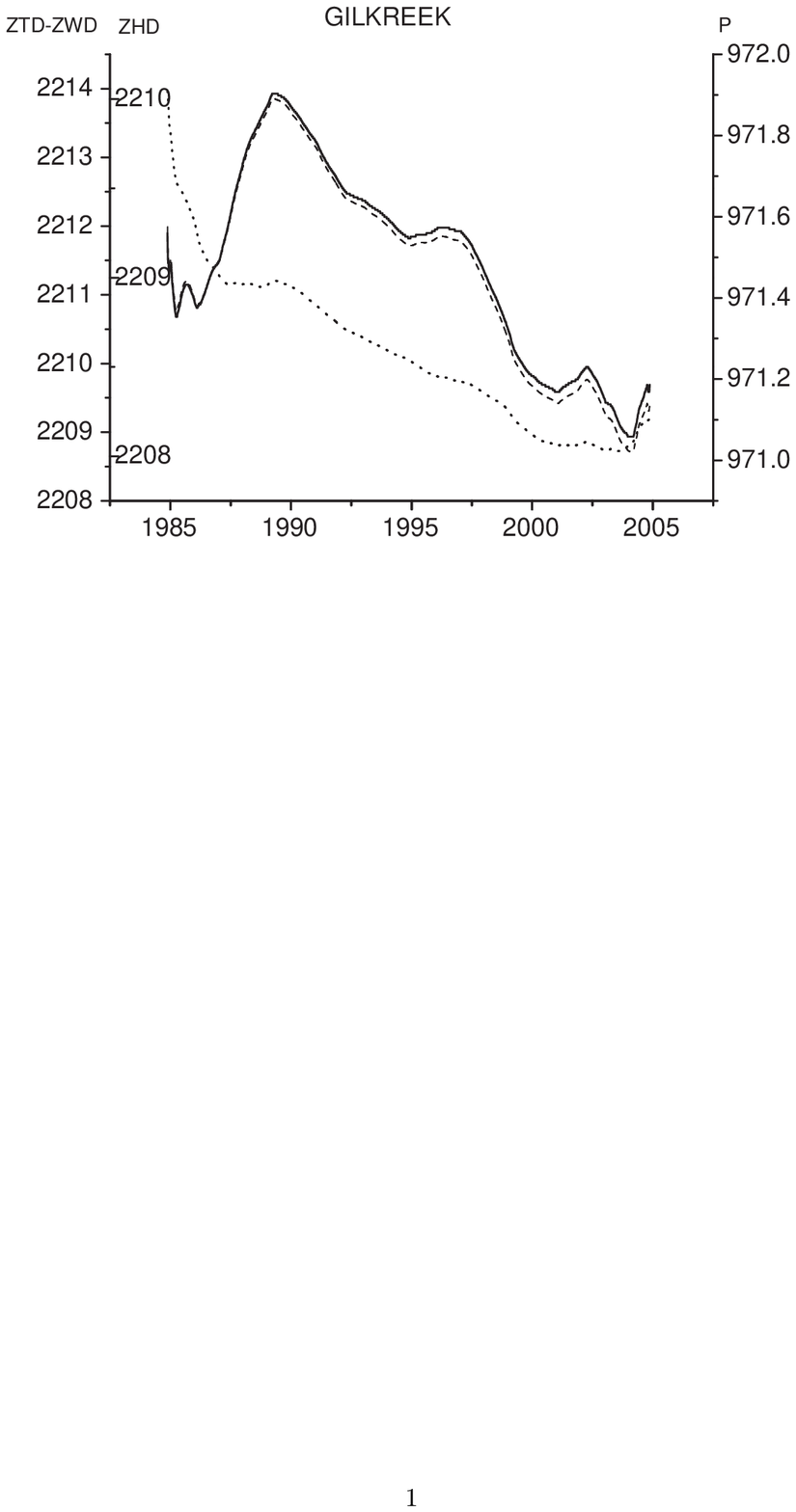}
\caption{SSA trends of hydrostatic zenith delay, pressure at the site and ZTD--ZWD.}
\label{fig:trendZHD}
\end{figure}

\begin{figure}
\centering
\epsfclipon \epsfxsize=0.45\textwidth \epsffile{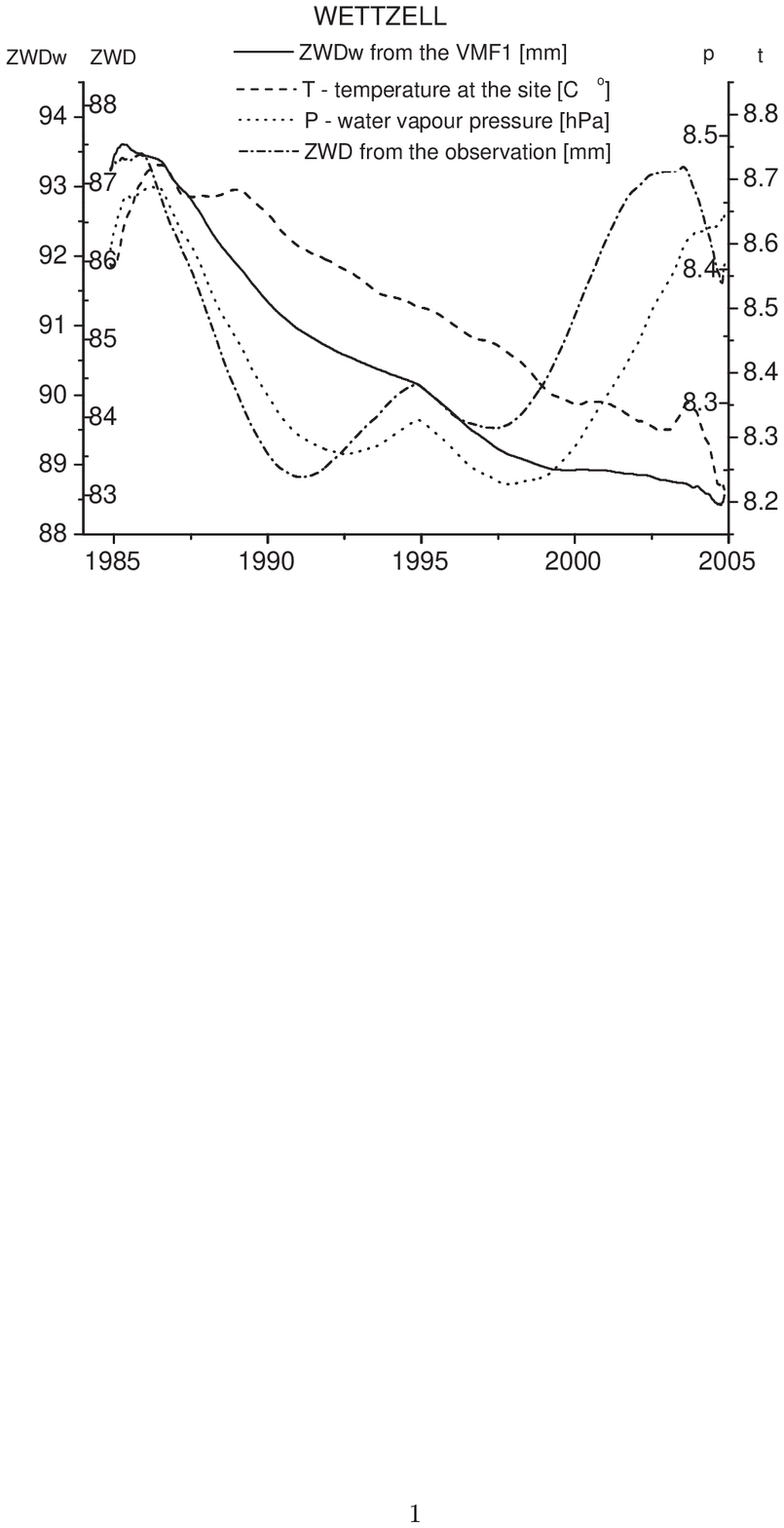}
\hspace{2em}
\epsfclipon \epsfxsize=0.45\textwidth \epsffile{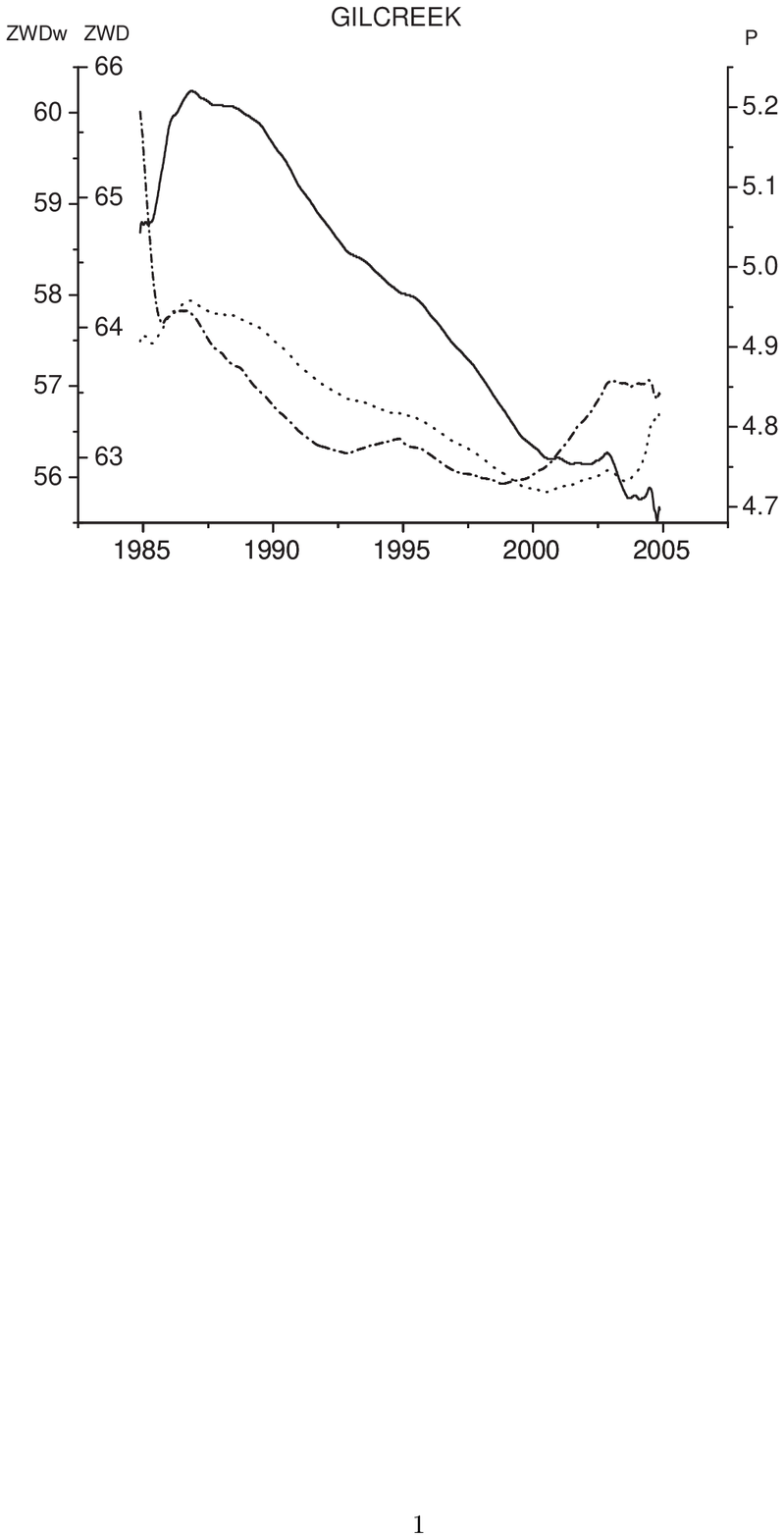}
\caption{SSA trends of wet zenith delay, temperature at the site, water vapor pressure and ZWD.}
\label{fig:trendZWD}
\end{figure}

It is interesting to compare non-linear trends found in the ZD with those
in the meteorological parameters. For this purpose, decomposition of
the meteorological parameters for Wettzell and Gilcreek has been
made using SSA.
Figure~\ref{fig:trendZHD} shows the trends for the hydrostatic zenith delay,
pressure at the site from the VMF1 and ZTD-ZWD computed from the IVS
combined series. The coincidence of all these curves is obvious for Wettzell.
For Gilcreek one can see the difference in shape of the curves.
Figure~\ref{fig:trendZWD} shows the trends for wet zenith delay,
temperature at the site, water vapor pressure at the site from the VMF1
and ZWD. The ZWD trend is very close to trend of water vapor pressure
for the Wettzell. But for the Gilcreek the similar curves
do not show such an agreement. The Gilcreek is the only station where we
failed to extract the trend of the site temperature series.

\begin{figure}
\centering
\epsfclipon \epsfxsize=\textwidth \epsffile{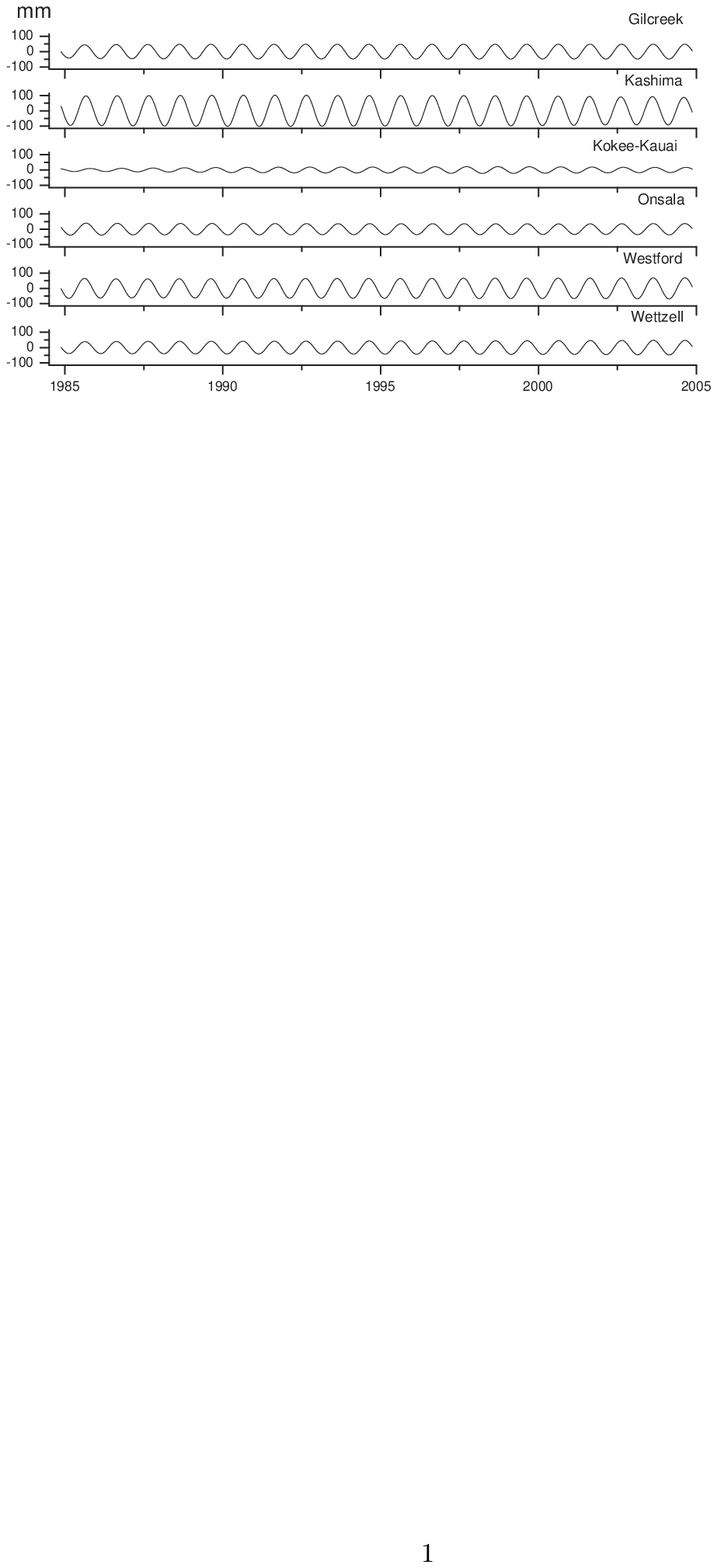}
\caption{The annual components in reconstructed ZWD series.}
\label{fig:annualZWD}
\end{figure}

\begin{figure}
\centering
\epsfclipon \epsfxsize=\textwidth \epsffile{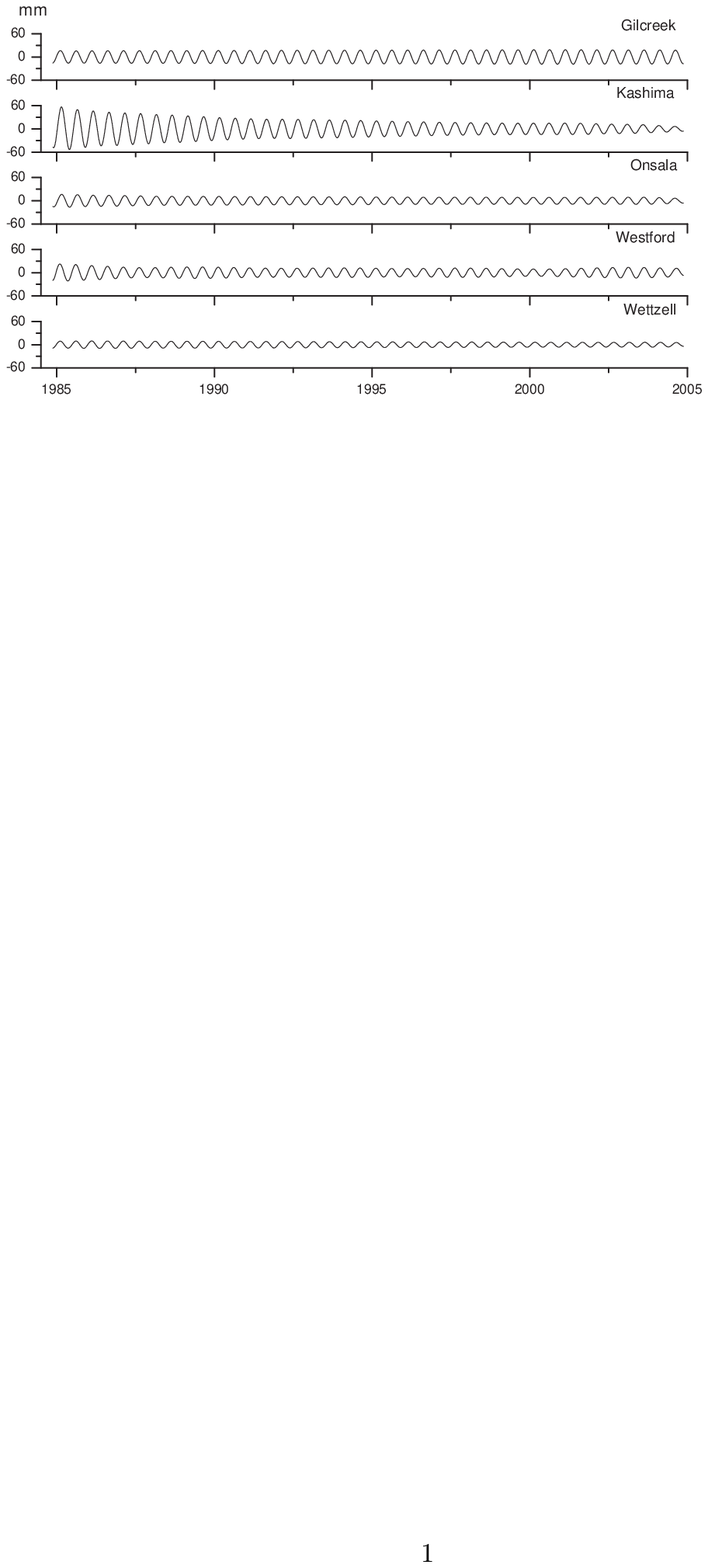}
\caption{The semiannual components in reconstructed ZWD series.}
\label{fig:semiannualZWD}
\end{figure}

Figures~\ref{fig:annualZWD} and~\ref{fig:semiannualZWD} show the annual
and semi-annual components ZWD for researched stations. The components have
a steady phase but variations of amplitude. It should be mentioned that the
contribution of the semiannual component is comparatively small, near noise
level, and therefore this component needs more careful study. For Kokee-Kauai
semiannual component was not found, and so for ZTD at Onsala.
For Kashima, the amplitude of the semiannual component is larger in the
beginning of the time interval than at the rest of interval.

\section{Conclusions}
In this paper, we have examined an ability of the SSA method in analysis
of the zenith troposphere delay. Non-linear trends and variations of the
amplitude of seasonal components have been detected. Some interesting
peculiarities in their behavior have individual character for every
stations of site. Comparison of the trends with meteorological parameters
also is presented to show possible similarities that deserve further
investigations.

\end{document}